\documentclass{article}
\usepackage{amsmath}
\usepackage{graphicx}
\usepackage{subfigure}
\usepackage{color}

\begin{document}

\title{\Large \bf ARE FOREST FIRES PREDICTABLE?}   

\author{
K.~Malarz$^*$,
S.~Kaczanowska,
K.~Ku{\l}akowski$^\dag$\\
\normalsize \em
Department of Theoretical and Computational Physics,
Faculty of Physics and Nuclear\\
\normalsize \em
Techniques,
University of Mining and Metallurgy (AGH)\\
\normalsize \em
al. Mickiewicza 30, PL-30059 Krak\'ow, Poland.\\
\normalsize \em
E-mail: $^*$malarz@agh.edu.pl, $^\dag$kulakowski@novell.ftj.agh.edu.pl
}

\maketitle

\begin{abstract}
\noindent
Dynamic mean field theory is applied to the problem of forest fires. The
starting point is the Monte Carlo simulation in a lattice of million cells. The
statistics of the clusters is obtained by means of the Hoshen--Kopelman
algorithm. We get the map $p_n\to p_{n+1}$, where $p_n$ is the probability
of finding a tree in a cell, and $n$ is the discrete time. We demonstrate
that the time evolution of $p$ is chaotic. The arguments are provided by the
calculation of the bifurcation diagram and the Lyapunov exponent. The
bifurcation diagram reveals several windows of stability, including periodic
orbits of length three, five and seven. For smaller lattices, the results of
the iteration are in qualitative agreement with the statistics of the forest
fires in Canada in years 1970--2000. \end{abstract}   

\section{Introduction}    
It would be of obvious interest if one could predict a forest fire. An
attractive aim would be also to evaluate the risk of the forest fires in a
given year, having the statistics of the fires in the past. Here we argue that
the time dependence of the forest area burned in a given year is inherently
chaotic. This means that such predictions are not possible in longer time
scale. The method used is the mean field equation based on an appropriate
cellular automaton.

Cellular automata (CA) is a modern tool for qualitative simulations of 
numerous problems in statistical mechanics and theory of dynamical 
systems \cite{wolfram,stauffer}. In general, the rules of CA are
probabilistic | they may depend on a random variable. When this dependence
is absent, the automaton is termed deterministic. These automata are most
exciting but least predictable. The behavior of a probabilistic cellular
automaton can be predicted successfully \cite{wooters} by means of a
nonlinear mean field equation. The solution is the time-dependent probability
distribution of cell states. The approach is termed ``mean field'', because the
correlations between the states of neighboring cells are lost within one time
step. However, they are lost also in the original probabilistic automaton
after several time steps. Similar technique can be applied also to investigate
deterministic CA \cite{wooters}, but in this case the results of the mean field
approximation are less reliable: the rules of an original deterministic
automaton can preserve the correlations forever.

Our aim here is to apply this mean field technique to the problem of the
forest fires. The list of references of the problem is vast and reveals its
numerous aspects, from purely computational to inherently biological ones
\cite{duarte}. In particular, some controversy has been mentioned \cite
{duarte} on the applicability of the percolation theory \cite{beer}, and on
the size dependence of results of possible experiments. The advantage of the
mean field approximation is that the finite-size fluctuations are eliminated.
Then, our results can be seen as a limit case both for simulations and
experiments. Below we demonstrate that the critical concentration for the
percolation problem is fairly reproduced by the forest fire simulation, if
the lattice is large enough.

The approach can be formulated as follows: The model forest is a
two-dimensional regular lattice. Each cell at the lattice may be occupied by
a tree with probability $p$. During a year (one time step), trees appear at
empty cells with a probability $rp(1-p)$, where $p$ is the previous
concentration of trees per cell, and $0\le r\le 1$ is a model control parameter.
This nonlinear character of the probability reflects the obvious tendency 
that a new tree appears less likely if there is almost no free space around 
the cell. This kind of dependence is known as the Verhulst term \cite{makino}.
All trees, those existing previously and those just grown, form clusters of
different sizes. A cluster is a set of trees which are connected by
nearest-neighbor bonds in the von Neumann neighborhood. Now, a probability
that a cluster is ignited is proportional to its size. Then, a fraction of
the forest burned each summer is just the mean square of the cluster size. We
will show that below the percolation threshold $p_c$, this fraction is
negligible. Above this concentration, most of the forest disappears | only
small clusters survive.

Summarizing our method, the time evolution equation for the concentration of
trees in the forest is a superposition of two maps, which describe a sequence
of growing (``spring'') and fire (``summer'') 
\begin{subequations}
\label{eq_map}
\begin{equation}
\label{eq_1}
p_{n+1}=W(p_n+rp_n(1-p_n)),
\end{equation} 
where the map $W(p)$ is
\begin{equation}
\label{eq_2}
W(p)=p-A(p)
\end{equation} 
\end{subequations} 
and $n$ enumerates years, i.e. the time steps. The function $A(p)$ is the
burned fraction of the total area of the forest, i.e. the mean square of the
cluster size. This fraction is evaluated by means of the Hoshen--Kopelman
algorithm \cite{hoshen}.

The purpose of this paper is to demonstrate, that the map $p_n\to p_{n+1}$
leads to chaotic behavior. We note that such a conclusion is not entirely
new. In 1990, Chen, Bak and Jensen \cite{chenbak} proposed a deterministic
coupled map lattice \cite{comment} for a model description of forest fires. In
this model, fire was introduced as a continuous process. The Lyapunov exponent
calculated there described the time evolution of the Hamming distance between
two nearby trajectories. Initial distance was set by a random perturbation,
added to each site at time $t_0$. As the distance was found to increase with
the power law, the value of the Lyapunov exponent was zero. Later on, this
model was generalized by Socolar, Grinstein and Jayaprakash \cite{socolar} by
adding a parameter, which allowed to get a positive Lyapunov exponent.

Both the above mentioned approach and this work have in common that they
are computational rather than oriented to experiment. However, within this
computational stream there are serious differences. The algorithm of Chen
et al. \cite{chenbak} is deterministic. The forest fire is continuous; a tree
burns spontaneously as soon as it is large enough. Our map can be treated as an
extrapolation of the output of a probabilistic automaton to the area of
infinite size. Then, a sequential iteration is applied, and the season of
growing and the season of fires are treated as two successive parts of one step
of the iteration. From Ref. \cite{duarte} we know that there are many
uncontrolled quantities relevant to the problem: moisture, humidity, landscape,
etc. As it is impossible to take them into account within a simple scheme, our
probabilistic approach seems to be closer to reality. Last but not least, our
approach gives a sequence of yearly burned areas which, although expressed in
arbitrary units, can be directly compared to available statistical data.

The paper is organized as follows: In Section \ref{secnum} we describe the
way how the Hoshen--Kopelman is used here. There, two different kinds of
statistics are assigned to casual (lightnings) and intentional (``human factor'') forest fires. In Section \ref{secres} the results on the Lyapunov exponent
and the bifurcation diagram are given. A remark is also added on the fire
size distribution. The comparison of our results to the statistical data on
forest fires in Canada in years 1970--2000 \cite{web} is included in Section
\ref{secexp}. The last section is devoted to discussion.

\section{Numerical approach} \label{secnum}
With the Hoshen--Kopelman algorithm \cite{hoshen} we are able to label all
occupied sites on a $L\times L$ large square lattice in such a way that the
sites with the same label belong to the same cluster and different labels are
associated to different clusters. In this way we get the cluster size
distribution, which allows us to obtain the function $A(p)$ and, in the next
step, to formulate the mean field map given in Eq. \eqref{eq_map}.

The beauty of the Hoshen--Kopelman algorithm for the cluster characterization
is that it goes through the lattice {\em only} once, and stores {\em only}
the current line of the length $L$. Thus it is particularly useful for the
bus-bar percolation problem investigation and it allows to check whether a
given occupied site at distance $x$ from an arbitrary chosen (usually the
first and full occupied) line is still connected to this line through the
occupied sites in lines $2, 3, \ldots, x-2, x-1$ \cite{vidales,aharony}.

\begin{figure}
\begin{center}
\includegraphics[width=5cm]{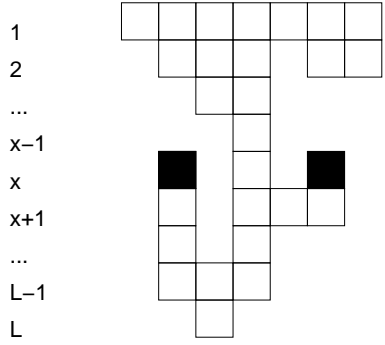}
\end{center}
\caption{Are black sites at the line $x$ recognized as a part of the cluster?
Algorithm goes from left to right, from top to bottom.}
\label{fig_clu}
\end{figure}

However, to construct the map \eqref{eq_2} it is necessary to evaluate the
mean cluster size (or its average mass) $L^2A(p)$. This requires an
information on all bonds to sites at the distance $x$ from the first line also
through the occupied sites at the lines $x+1, x+2, \ldots, L-1, L$, as it is
presented on Fig. \ref{fig_clu}.
This in turn requires storing of
the whole lattice for all the time of simulation and passing through the
lattice twice | what wastes computer memory and machine time, but ensures
proper labeling of the clusters.

The simulation is carried out on $10^3\times 10^3$ square lattice,
with a fraction of $p+rp(1-p)$ occupied sites for different values of the
initial trees concentration $0\le p_0\le 1$. An average concentration of trees
in a ``green'' (but soon ``red'' and consequently ``black'') cluster
\[
A(p)=\dfrac{1}{L^2}\sum_i w_i s_i
\]
may be evaluated with two kinds of
weights
\[
w'_i=s_i/L^2 \quad \text{ and } \quad w''_i=s_i/\sum_is_i,
\]
which describe the probability of finding a cluster of size $s_i$
on the lattice and the probability of choosing an $i$-th cluster among all
other clusters, respectively.
These statistics can be interpreted as reflecting
the probability of a casual fire started from a lightning, which can strike
at an empty cell ($w'$), or of a fire started intentionally at a randomly
selected tree ($w''$). The obtained map \eqref{eq_map} for each $p$ is the
average over a hundred of independent lattice realizations.

Below we concentrate on the case of the fires of natural origin. The results
for the other case will be only listed. The details of the latter calculation
will be discussed more thoroughly elsewhere.

\section{Results}
\label{secres}  
\subsection{The map}  
\begin{figure}[tbp]
\begin{center}
\includegraphics[angle=-90,width=0.9\textwidth]{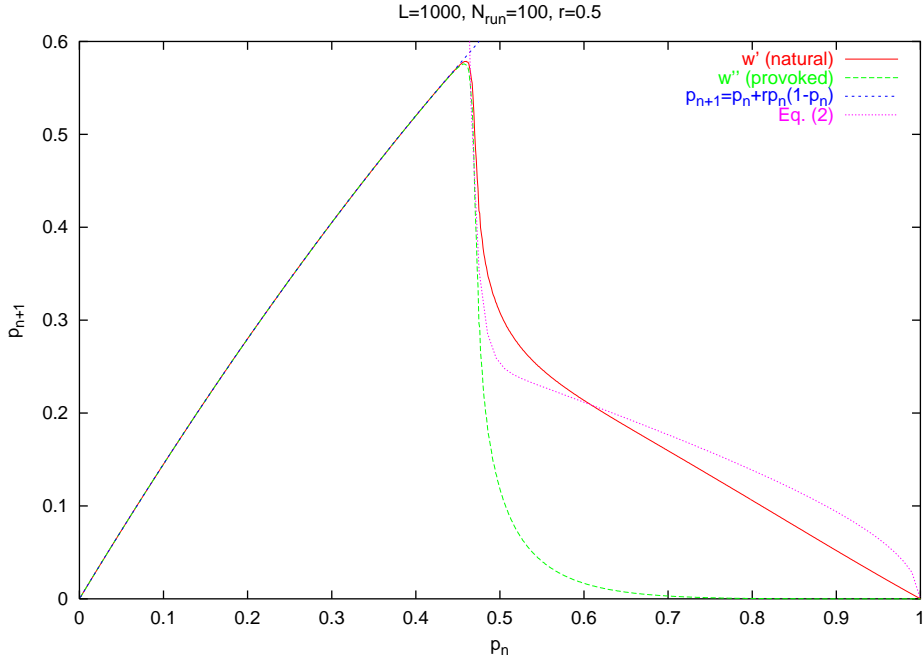}
\end{center}
\caption{Map for model with average burned area $A(p)$ evaluated for different
weights $w'$ (natural) and $w''$ (provoked). Before percolation
threshold burned areas are negligible and the map is given by argument of
function $W$ in Eq. \eqref{eq_1}. Also map for recipe \eqref{eq_fit} is
included.}
\label{fig_map}
\end{figure}

The map $p_n \to p_{n+1}$ is given in Fig. \ref{fig_map}.
Two plots refer to two kinds of statistics, described in Section \ref{secnum}.
We see that the plots are the same below
the maximum of the curve, which is near the percolation limit for the square
lattice $p_c=0.59273$ \cite{stauffer,stakuos}. Actually, below this limit both plots
are identical to the curve $p+rp(1-p)$ from Eq. \eqref{eq_1}. This means,
that for $p<p_c$ the fires are negligible. Above the percolation limit, the
curves remain similar for $p$ close to $p_c$, but for higher $p$ the curve
for the weight $w'$ is remarkably higher, than the curve for the weight
$w''$. It is clear that intentional fires ($w''$) are more devastating,
because in this case it is sure that a tree is ignited.

\subsection{The bifurcation diagram} 
\begin{figure}
\begin{center}
\subfigure{
\label{fig_bif_a}
(a) \includegraphics[angle=-90,width=0.9\textwidth]{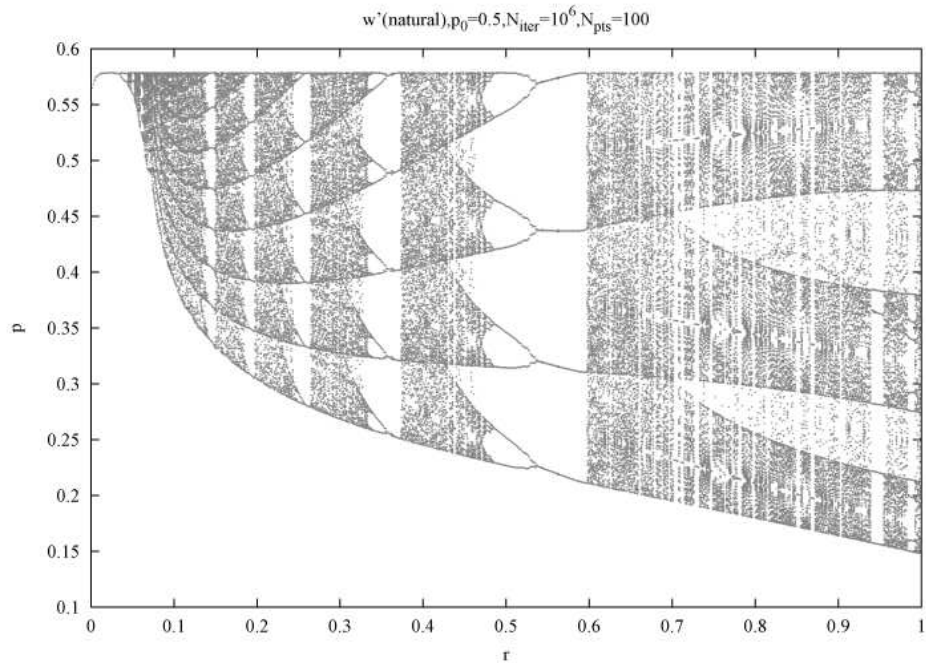}
}
\subfigure{
\label{fig_bif_b}
(b) \includegraphics[angle=-90,width=0.9\textwidth]{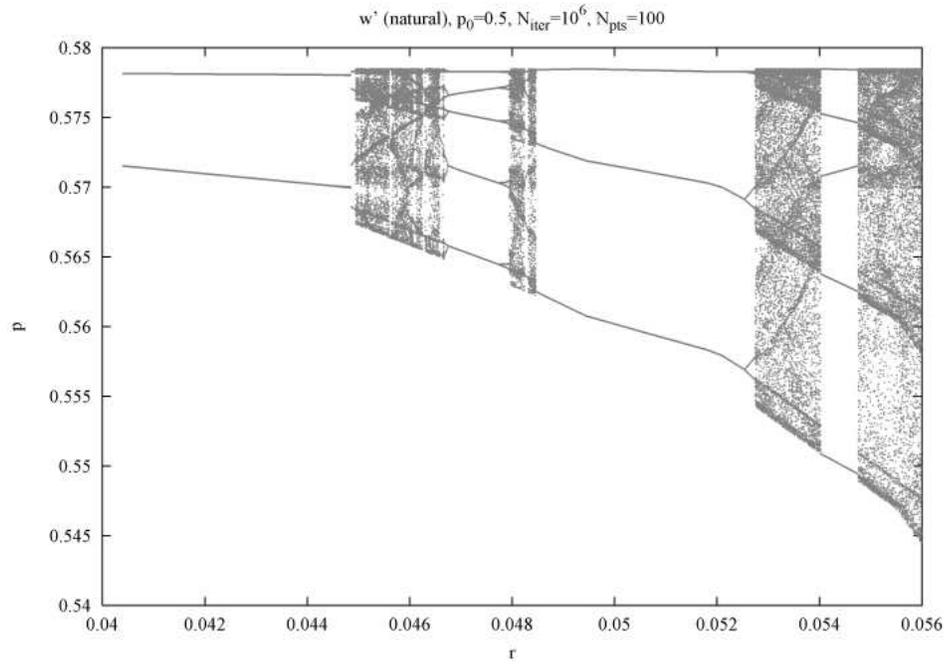}
}
\end{center}
\caption{Bifurcation diagram for weights (a)--(b) $w'$ (natural).
The last hundred values of $p$ for each $r$ is presented.
Also diagram for (c) map given by Eq. \eqref{eq_fit} is included.}
\label{fig_bif}
\end{figure}
\addtocounter{figure}{-1}
\begin{figure}
\addtocounter{subfigure}{2}
\subfigure{
(c) \label{fig_bif_c}
\includegraphics[angle=-90,width=0.9\textwidth]{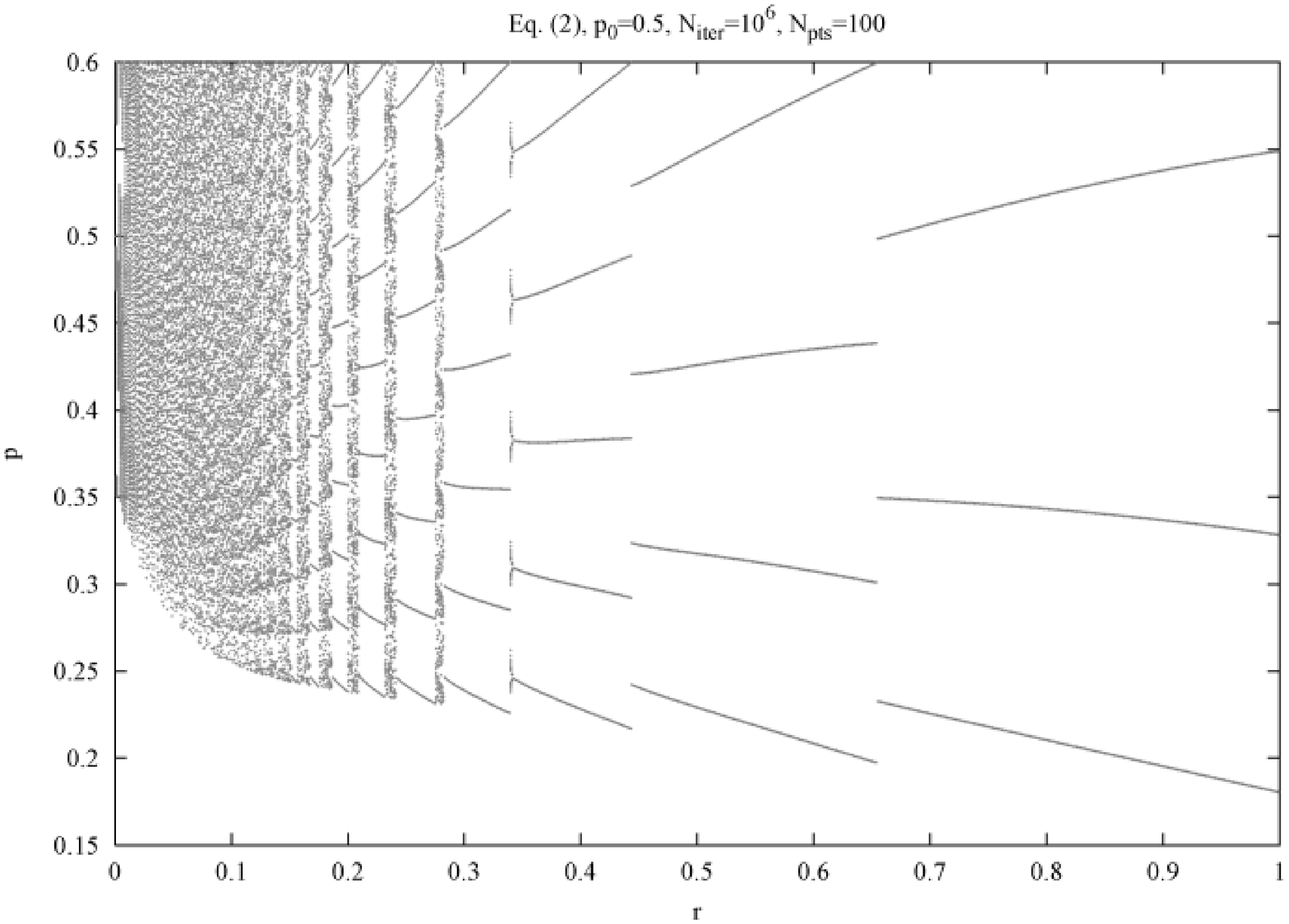}
}
\caption{Continued from page \pageref{fig_bif}...}
\end{figure}
The rest of the results refer to the case of casual fires, i.e. the weight
$w'$. In Fig. \ref{fig_bif} we show the bifurcation diagram. This diagram is obtained
with the map $p_n\to p_{n+1}$, supplemented by straight lines joining
neighboring $10^3$ numerical points. We can see the windows of stable cycles of
lengths four, eight, five, ten, six, twelve, seven and eight again (Fig. \ref{fig_bif_a}),
five, three, five and two in Fig. \ref{fig_bif_b},
when the parameter $r$ decreases. The presence of the cycle three proves that,
by means of the Sharkovskii theorem, all possible lengths of the limit cycles
are present \cite{schuster}.

A question arises, if our bifurcation diagram is not an artificial consequence
of the finite grid of the map and/or the finiteness of the lattice. To check
this point, we calculated also the bifurcation diagram for a piecewisely
analytical map $G(H(p))$, where 
\begin{equation}
\label{eq_fit}
\begin{split}
H(p) & = p+rp(1-p),\\
G(p) & =
\begin{cases}
0.1\exp[100(0.6-p)]+0.4\sqrt{1-p} & \iff p>0.6,\\
p & \iff p\le 0.6.
\end{cases}
\end{split}
\end{equation}
This map is also shown in Fig. \ref{fig_map}.
It is designed not as to fit the plot
obtained within the Hoshen--Kopelman algorithm, but rather to investigate the
consequences of the particular shape of this plot. The obtained bifurcation
diagram (Fig. \ref{fig_bif_c}) shows the windows of stability, with the
width which increases with the parameter $r$. The lengths of the cycles
decreases with $r$ in the same way as presented in Fig. \ref{fig_bif_a}. We
deduce that this part of the bifurcation diagram is generic.

To each cycle, an admissible word can be assigned according to the rules of
symbolic dynamics. We have only to check the values of $p$ in the sequence
of a given limit cycle. The rule is as follows: The value of $p$ preceding
the largest $p$ (close to $p_c$) in the cycle is labeled $C$. Let us call this
value $p^*$. The next value (close to $p_c$) is labeled $R$. Any other $p$ is
labeled $R$ if it is larger than $p^*$, and it is labeled $L$ if its value is
smaller than $p^*$ \cite{haobailin}. We use this rule to assign symbolic words
to each cycle in Fig. \ref{fig_bif_a}, and we compared their sequence to the
standard one for the unimodal maps \cite{metro}. The sequences of the cycles
presented in Figs. \ref{fig_bif_a} and \ref{fig_bif_b} are respectively
$RL^2$, $RL^3RL^2$, $RL^3$, $RL^4RL^3$, $RL^4$, $RL^5RL^4$, $RL^5$, $RL^6$
and $RL^2R$, $RL$, $RLR^2$, $R$ with decreasing parameter $r$. Note that $C$
is omitted for convenience at the beginning of each word. We have found that
these sequences indicate opposite directions of the increase of the parameter
$r$. According to the results discussed in the preceding paragraph, the left
part of the bifurcation diagram (Fig. \ref{fig_bif_b}) is supposed to
vanish in the thermodynamic limit. Still, the conclusion on all possible
lengths of the limit cycles holds, as it follows from the presence of cycle
five (all lengths but three) and cycle seven (all but three and five)
\cite{schuster}. For the bifurcation diagram in Fig. \ref{fig_bif_c} the
above prescription gives words $RL^n$. However, in this case the maximum of
the map does not appear in the limit cycle, except at the bifurcation points
where the cycle length changes. In this sense the letter $C$ should not be
used.

We have checked that the bifurcation diagram for weigths $w''$
seems to be a homogeneous spot. Still, we have found the cycle of length three by the symbolic dynamics method \cite{haobailin}. The details will be given elsewhere.

\subsection{The Lyapunov exponent}
\begin{figure}
\begin{center}
\includegraphics[angle=-90,width=0.9\textwidth]{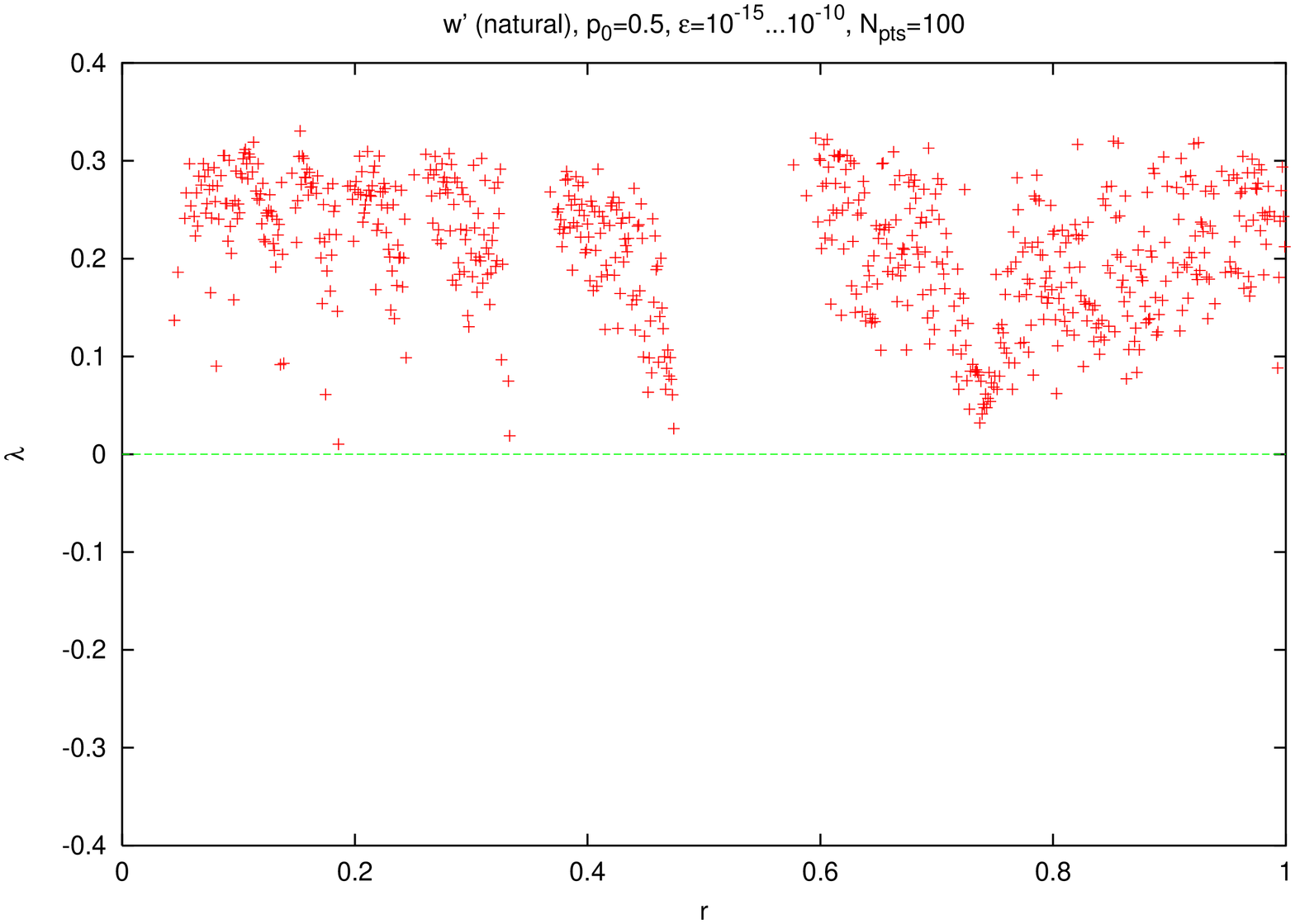}
\end{center}
\caption{Positive Lyapunov exponent for different model control parameter $r$.
Results of evaluating $\lambda$ are averaged over few differences
$\varepsilon$ between two values of initial concentration $p_0$. The linear
fit to the first hundred points of the Fig. \ref{fig_log_b} were used.}
\label{fig_lya}
\end{figure}

\begin{figure}
\begin{center}
\subfigure{
\label{fig_log_a}
(a) \includegraphics[angle=-90,width=0.9\textwidth]{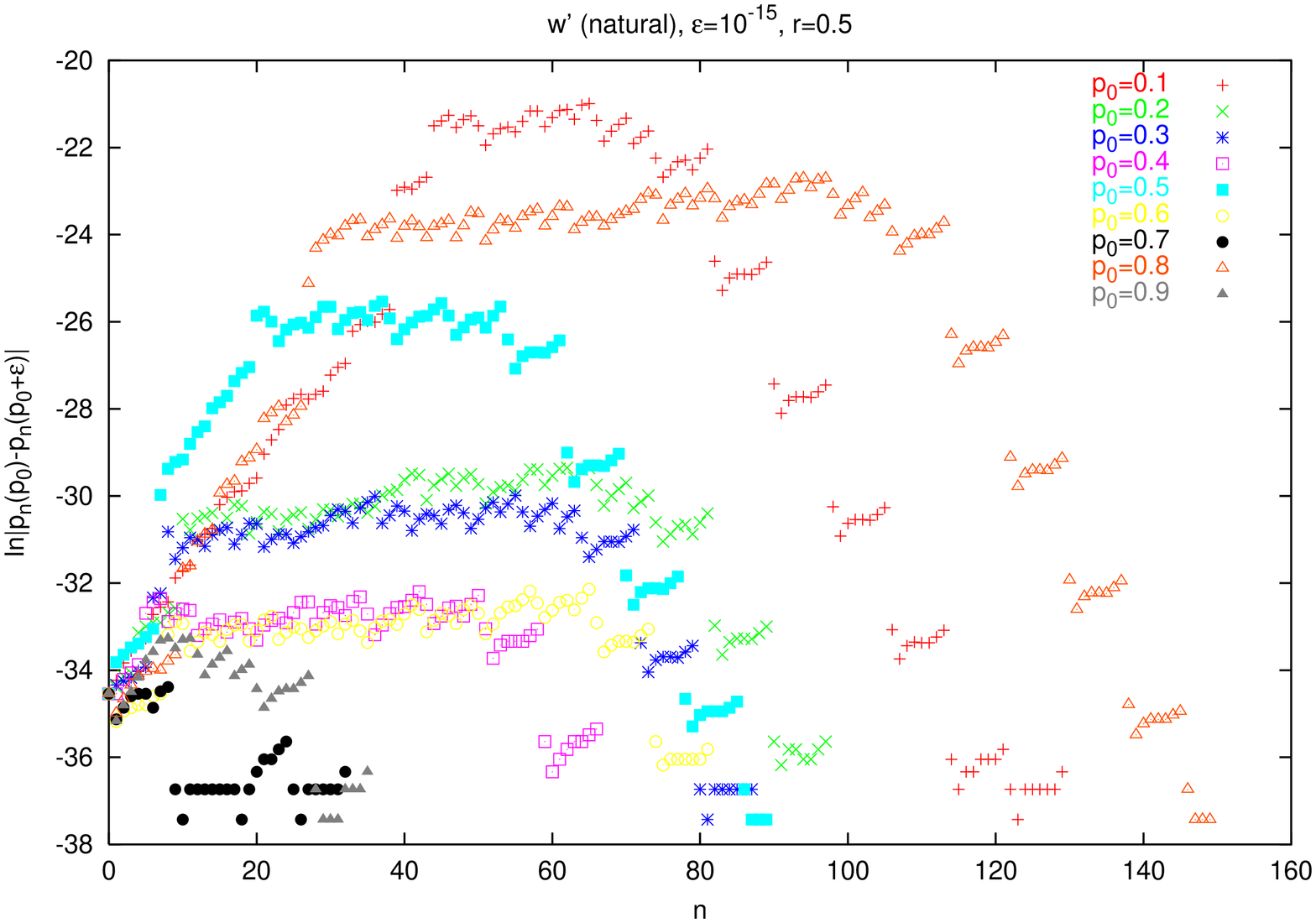}
}
\subfigure{
\label{fig_log_b}
(b) \includegraphics[angle=-90,width=0.9\textwidth]{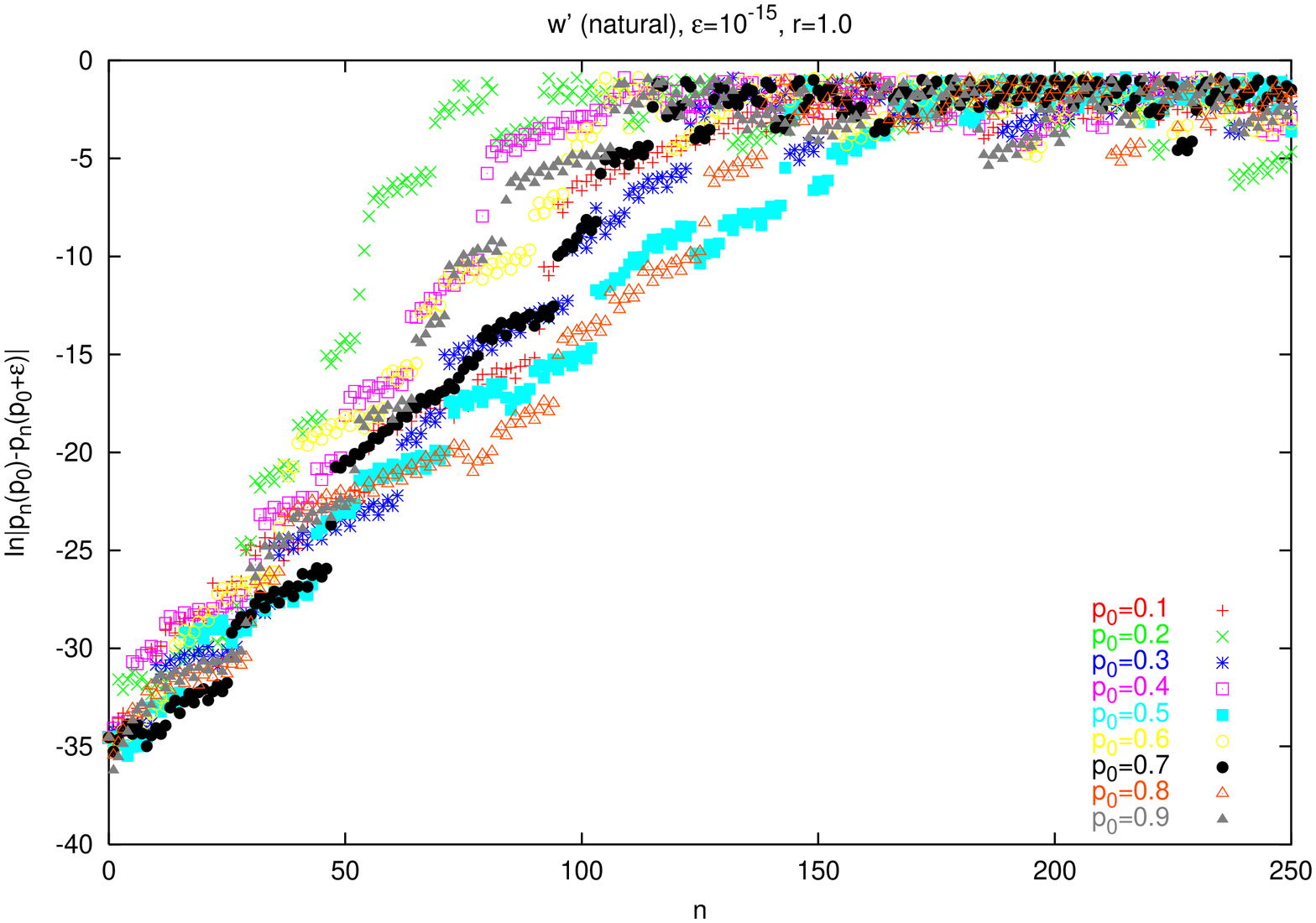}
}
\end{center}
\caption{Times dependence of the difference between two trajectories which
start from $p_0$ and $p_0+\varepsilon$.
Here $\varepsilon=10^{-15}$ and (a) $r=0.5$, (b) $r=1.0$.}
\label{fig_log}
\end{figure}

In Fig. \ref{fig_lya} we show the largest Lyapunov exponent as dependent on
the parameter $r$. Only positive values are shown, and the spaces between
data agree with the windows of stability in the bifurcation diagram. We have
found that the negative values of the Lyapunov exponent are much more
difficult to be calculated when we have a limit cycle and not a fixed point.
In this case, the time dependence of the difference between trajectories in
not monotonic. A trajectory can be trapped by a limit cycle with a shifted
phase, which leads to a fixing of the difference between trajectories even in
the case of a stable limit cycle.  Moreover, sometimes this trapping can be
only temporal. As the result, in the presence of a stable limit cycle the
difference between trajectories can depend on the initial point. These
complexities are revealed in Fig. \ref{fig_log}.

The Lyapunov exponent for weights $w''$ seems to be
positive in almost the whole range of the parameter $r$. This confirms that the
windows of stability are much more narrow.

\subsection{Fire size distribution}
\begin{figure}
\begin{center}
\subfigure{
\label{fig_hist_a}
(a) \includegraphics[angle=-90,width=0.9\textwidth]{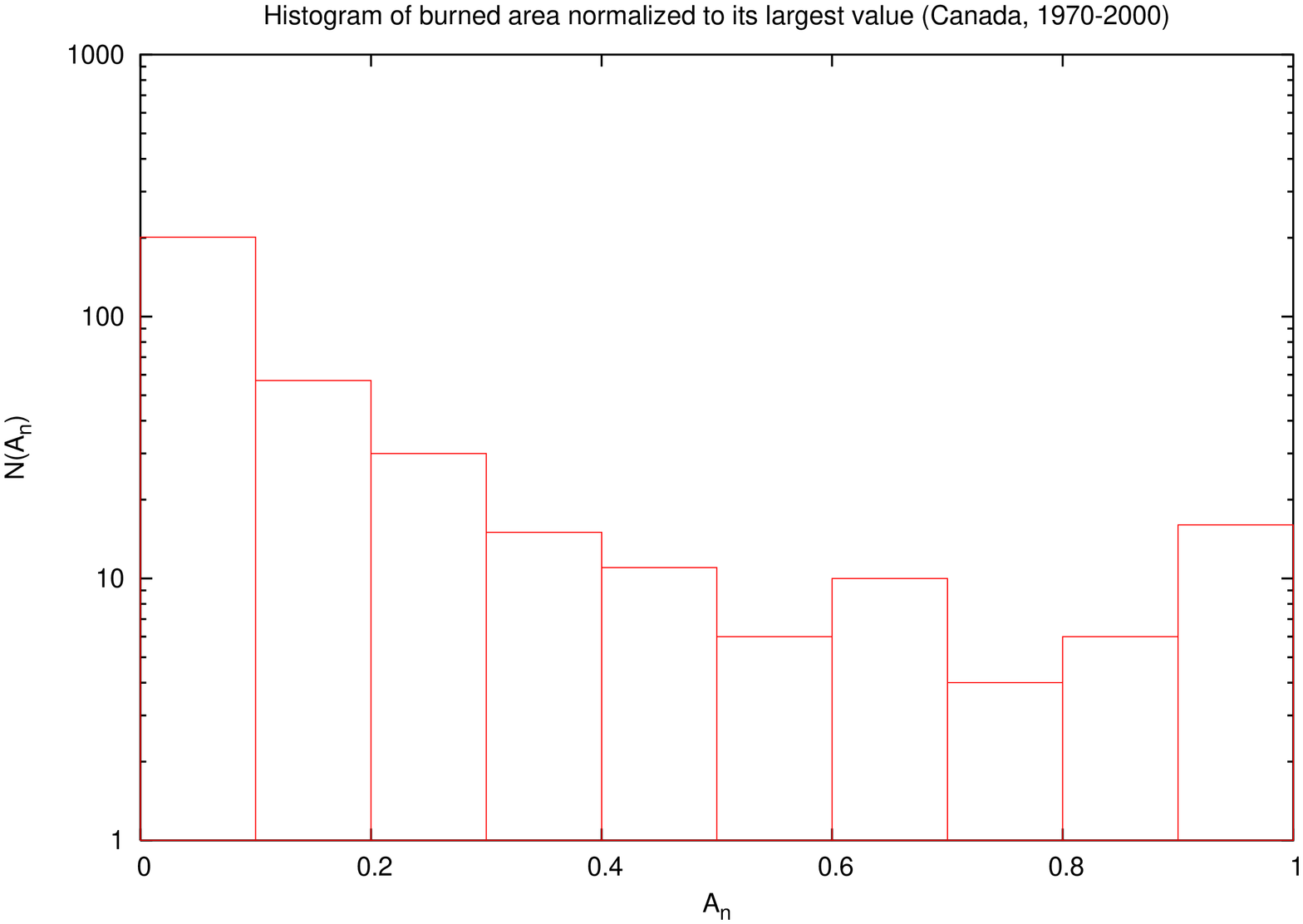}
}
\subfigure{
\label{fig_hist_b}
(b) \includegraphics[angle=-90,width=0.9\textwidth]{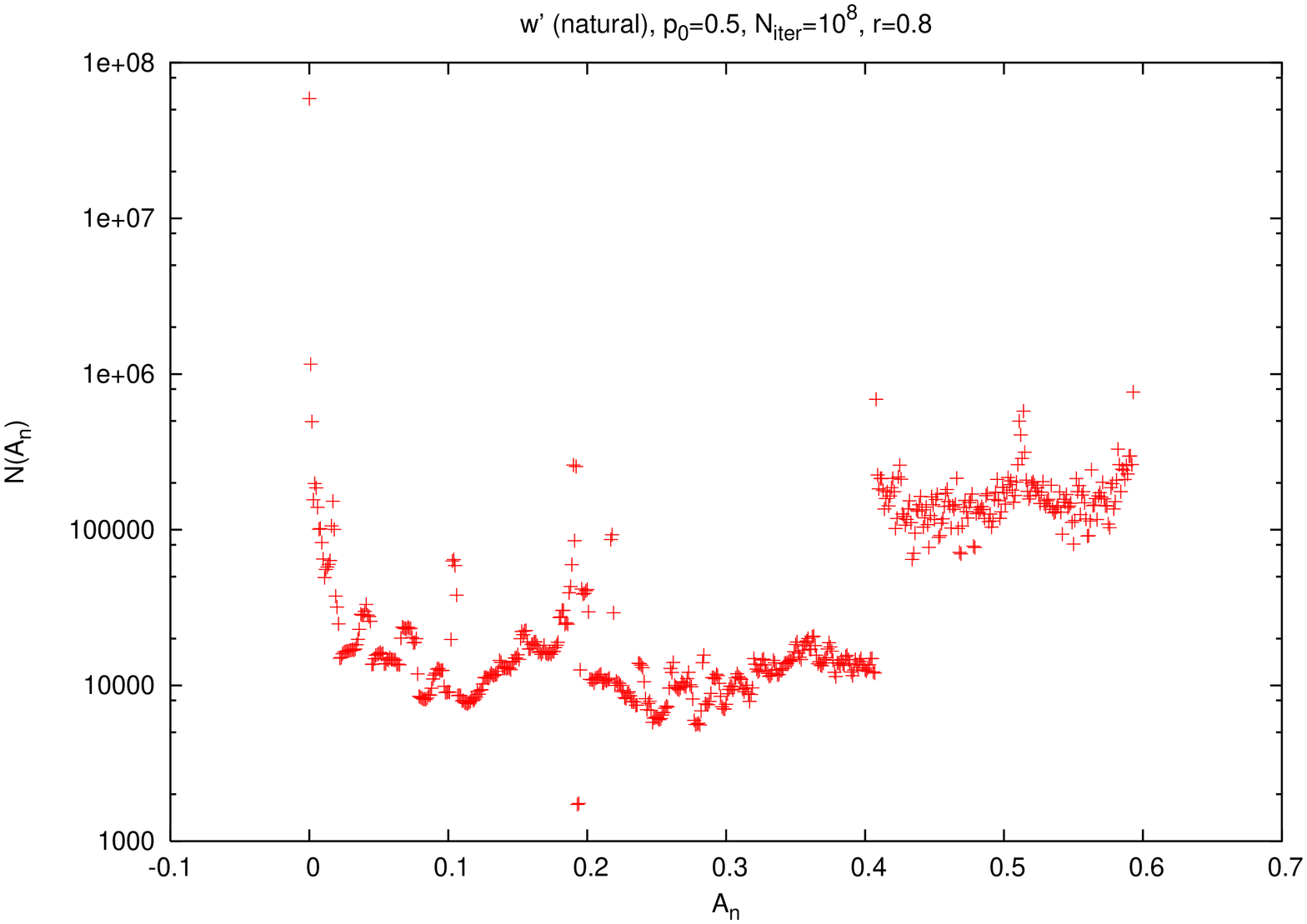}
}
\end{center}
\caption{Histogram of burned area (a) in Canada, 1970--2000 \cite{web}
rescaled to the largest fire in a given territory, and (b) obtained by means of computer simulation.}
\label{fig_hist}
\end{figure}

We have calculated the histogram of the size of fires, given by the function
$A(p_n)\equiv A_n$. Typical results for the map calculated by Eq.
\eqref{eq_map} for the
weights $w'$ are given in Fig. \ref{fig_hist_b}. As we see, the ``fires of
size zero'' are most
frequent. However, larger fires can be more likely that smaller ones. This kind
of curves do not show any self-similarity.
\begin{figure}
\begin{center}
\includegraphics[angle=-90,width=0.9\textwidth]{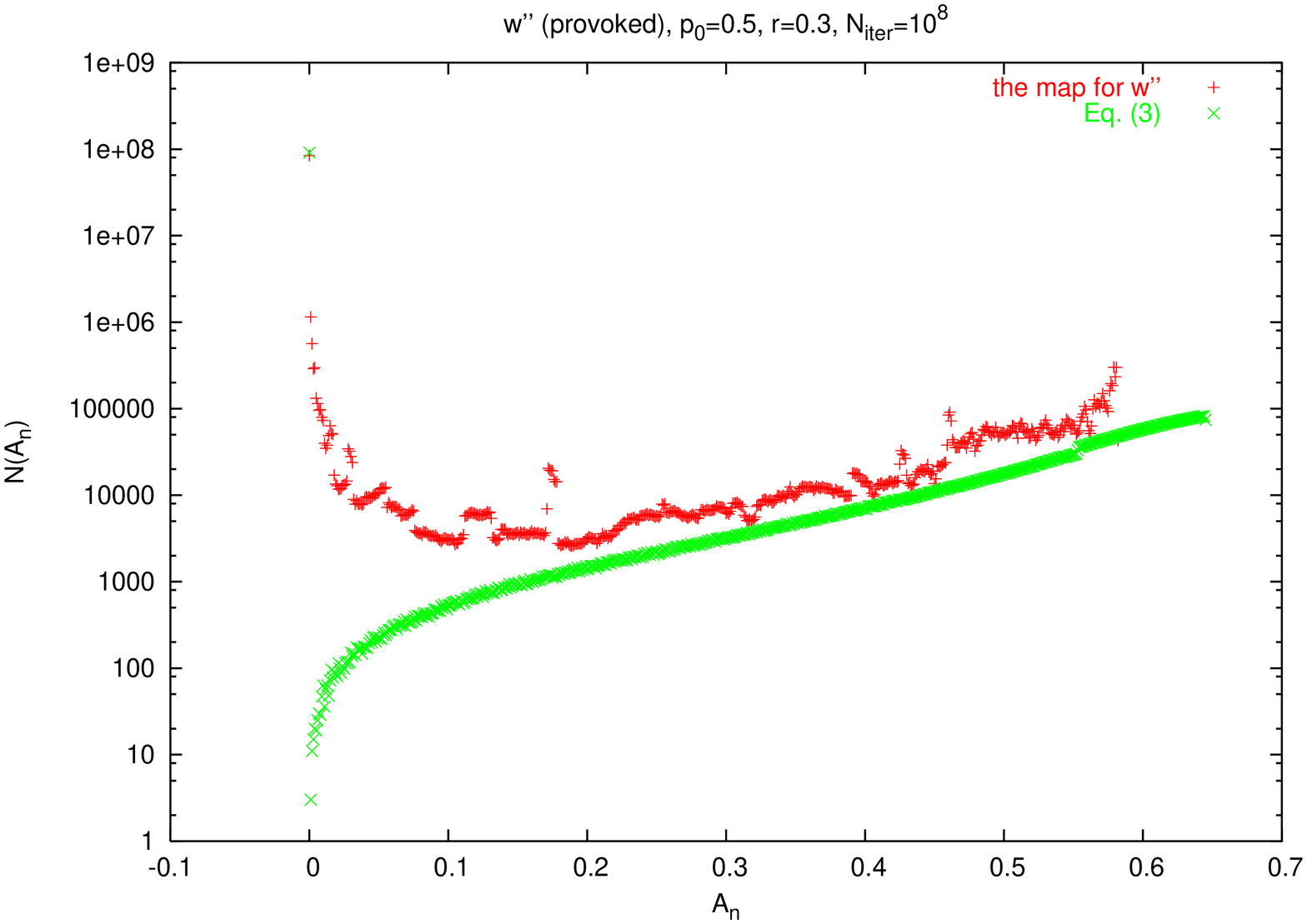}
\end{center}
\caption{Histogram of burned area for weights $w''$.}
\label{fig_hist_wbis}
\end{figure}

A serious difference between the results for $w'$ and $w''$ is found for the
histogram of the forest fires. An example for $r=0.3$ is presented in Fig.
\ref{fig_hist_wbis}.
For large fires, the curve seems to follow the exponential dependence. 
To capture this tendency, we have fitted the map for $w''$ by a curve similar
to Eq. \eqref{eq_fit}, where the function $G(p)$ is chosen as 
\begin{equation}
G(p)=p_c \exp (-12.1735 \sqrt{p-p_c})
\end{equation}
if $p>p_c=0.59237$.
The obtained histogram is added to Fig. \ref{fig_hist_wbis}. Note that the
latter curve is monotonic except the point at $A=0$, i.e. the events without a
fire. As we see, small finite fires are neglected by the above analytic
approximation.

\section{Comparison with experimental data}
\label{secexp} 
Our map $p_n \to p_{n+1}$ is obtained for the lattice of million sites, as an
average over hundred realizations. The accessible statistical data \cite{web}
relate to areas larger than million trees each, but to thirty ``iterations''
for one initial state of the forest. On the other hand, several uncontrolled
factors (moisture, etc.) cannot be taken into account within our simplified
approach. We expect that these agents lead to some noise, which is absent in
our deterministic map. This noise can be reproduced here by decreasing the
model lattice. That is why we are going to compare the experimental data
\cite{web} with the results of our simulation performed for a small lattice.

\begin{figure}
\begin{center}
\subfigure{
\label{fig_exp_a}
(a) \includegraphics[angle=-90,width=0.9\textwidth]{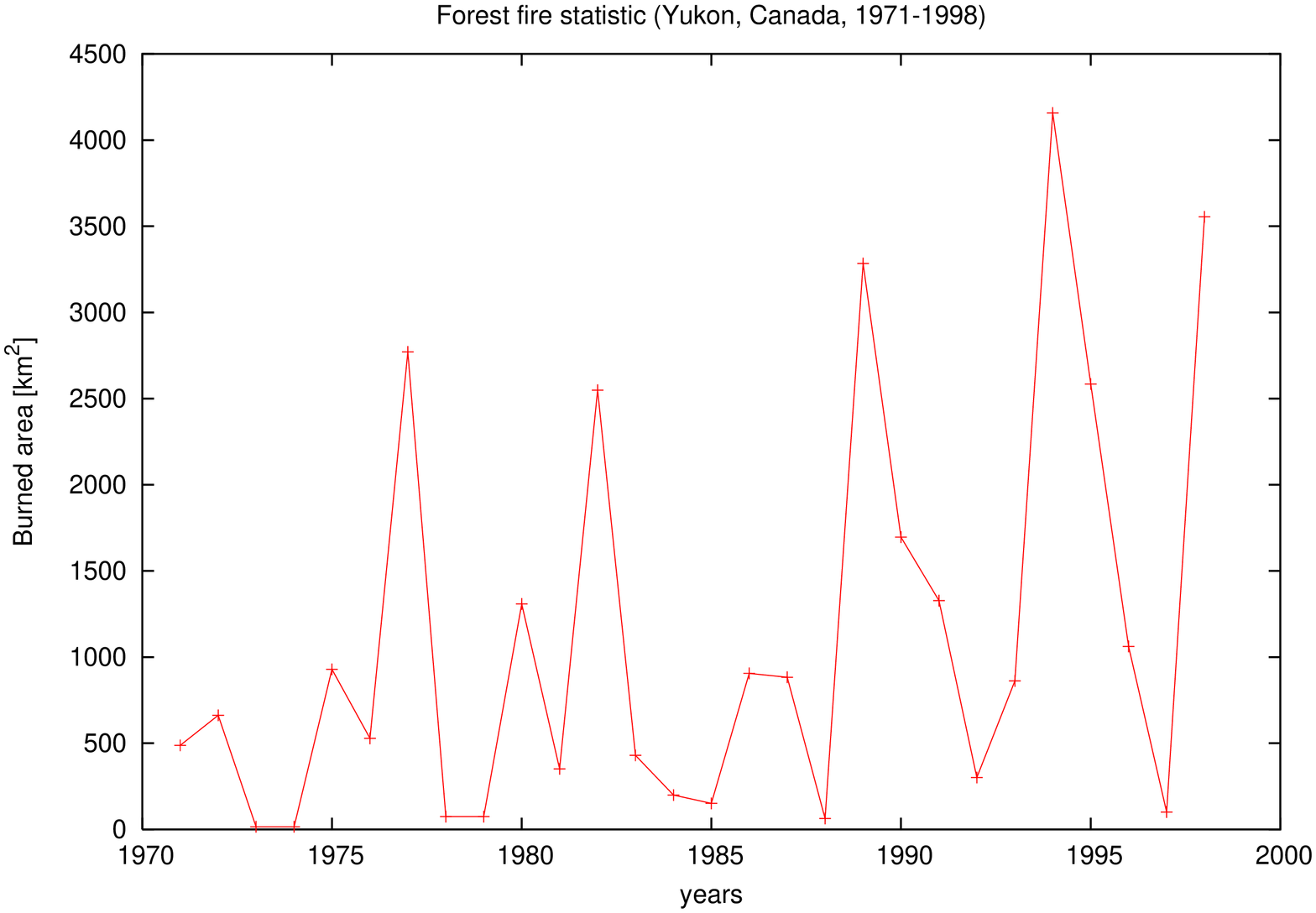}
}
\subfigure{
\label{fig_exp_b}
(b) \includegraphics[angle=-90,width=0.9\textwidth]{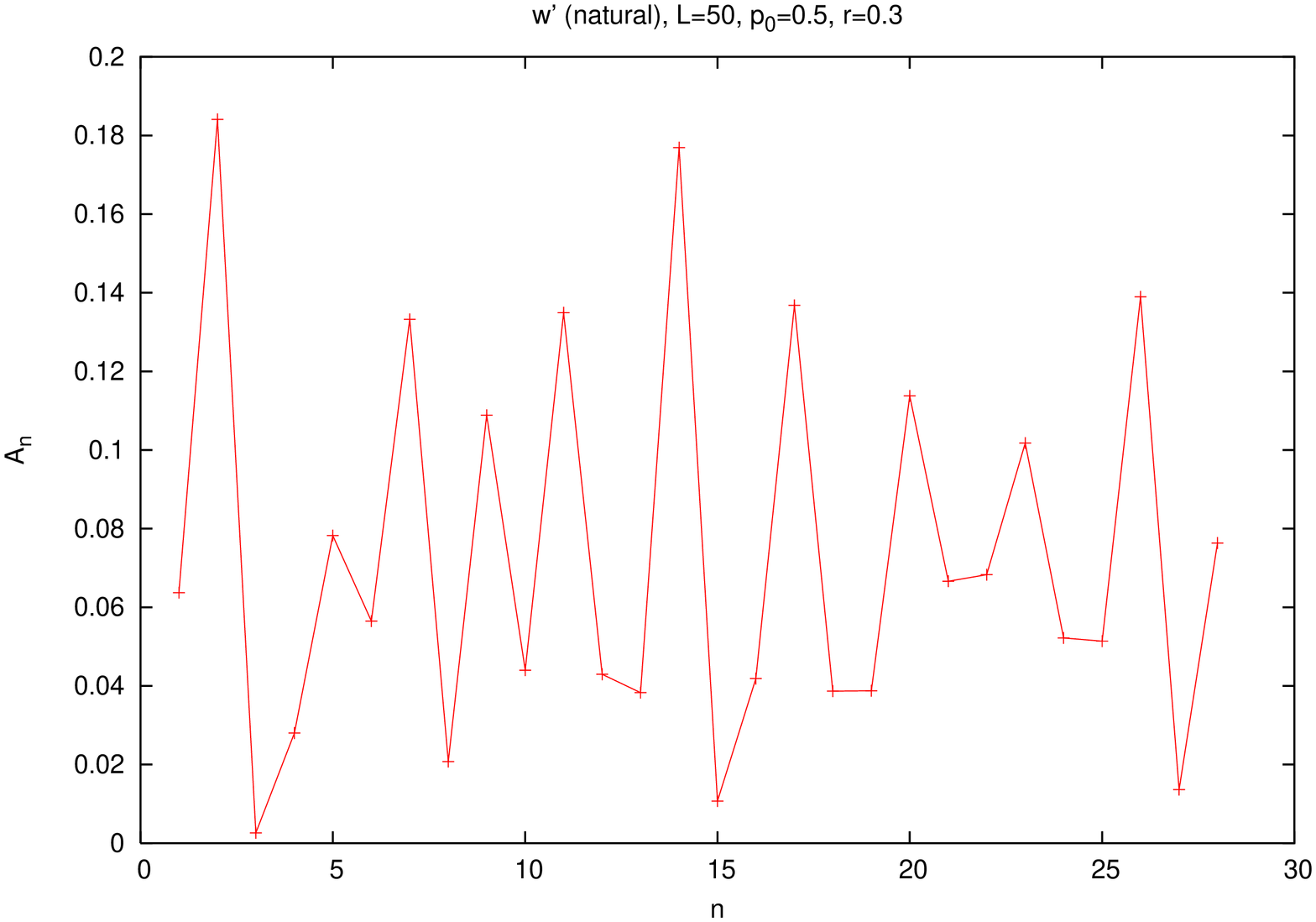}
}
\end{center}
\caption{Forest fire statistic for (a) real and (b) computer experiment.}
\label{fig_exp}
\end{figure}

We have checked that the time dependence of the size of the fires of both
experiment and calculations show the same kind of oscillations. Examples are
given in Fig. \ref{fig_exp}. In Fig. \ref{fig_exp_a} the data are
presented on the forest fire in Yukon, Canada, in years 1971--1998
\cite{web}.
In Fig. \ref{fig_exp_b} the same data are shown for the simulation
performed on the lattice of $50\times 50$ cells. As we see, the qualitative
features of the data are reflected by the simulation.

We also checked that the data on other provinces in Canada in years
1970--2000 \cite{web} do not differ in any essential way. In Fig.
\ref{fig_hist_a} we show a
collective histogram for experimental data. For each province or territory,
the fires are rescaled to the maximal fire in this region within the years
1970--2000. The statistics does not allow to speculate on the size distribution
of large fires. However, it is obvious that very small fires are most
frequent. This particular effect is reproduced in our numerical results
(Fig. \ref{fig_hist_b} and \ref{fig_hist_wbis}).

\section{Discussion}
\label{seclast}
The bifurcation diagram for weights $w'$ in Figs.
\ref{fig_bif_a}--\ref{fig_bif_b}
reveals an unusual structure, which is a conglomerate of two diagrams. On the
right side of the plot we observe some period doubling when the parameter $r$
decreases. On the left, clear period doublings are observed from a fixed
point to a limit cycle of length two, and then to a cycle of length four,
when the parameter $r$ increases. This left part of the bifurcation diagram is
due to the size effect of finite lattice. On the other hand, the map for
weights $w''$ constructed in the same way does not reveal wide windows of
stability. This means that the period doublings and the windows are sensitive
to the shape of the map on the right side of the percolation threshold. We
could only prove that the cycles of all lengths are present there, by means of
the presence of the cycle of period three. The map is similar to the tent map,
where no windows are observed.

Concluding this point, the bifurcation diagram reveals that the map is not
unimodal, what is not a surprise. Its structure may be caused by an
intermittent character of the trajectories for small $r$. In this case, if
the initial value of $p$ is small, forest fires do not occur in several years
and the evolution is governed by its parabolic part $p+rp(1-p)$, which
remains almost constant through a long time. The diagrams in Figs.
\ref{fig_bif_a} and \ref{fig_bif_c} are
similar to the diagram obtained in \cite{budd}. It was produced using a
discontinuous map, where the time evolution was also intermittent. On
the other hand, for larger $r$ the flat shape of the map above $p_c$ stabilizes
the limit cycles. We note that the intermittent character of the problem is to
some extent reflected also by the experimental data, where the majority of
fires is of small sizes.

The histogram in Fig. \ref{fig_hist_wbis} for weights $w''$ reveals an exponential curve of the
number of fires of size $A$ in the range of intermediate and large $A$.
The number of fires of size $A$ increases with $A$. This result is
in contradiction to what is known from literature on the self-organized
criticality \cite{drossel}. We interpret it as a particular consequence of
our assumption that fires are initiated more likely in large clusters. In the
approach presented in \cite{drossel}, a tree is ignited spontaneously if it is
old enough.

Coming back to the problem of predictability, we note that when a
deterministic mean field theory in the thermodynamic limit indicates the
chaotic behavior of the system, it is hard to imagine that any other approach
can produce a controllable solution. The exception is the the presence of the
windows of stability, which allows us to predict the size of a forest fire
for given values of $r$. Still, each limit cycle contains large fires,
indicated by the letter $R$ in the respective word. Obviously, this kind of
trajectory is not recommended by forest rangers. Concluding, both the
bifurcation diagram and the Lyapunov exponent reveal that chaos is indeed
present in the time evolution of the forest fire.

It seems also that the mean field approach can be a reference method in
investigating other problems where probabilistic cellular automata are useful.
A list of such problems includes immunology \cite{zorzenon} and magnetic
systems \cite{acharyya}. To apply a map or a differential equation? The answer
must be specific for a given problem.

\section*{Acknowledgments}
The authors thank to Professor Dietrich Stauffer for valuable comments.
The simulations were carried out in ACK-CYFRONET-AGH. The
time on SGI 2800 machine is financed by the Polish Committee for Scientific
Research (KBN) with grant No. KBN/\-SGI2800/\-022/\-2002.



\begin{thebibliography}{99}
\bibitem{comment} The authors of \cite{chenbak} refer to their technique as
to a cellular automaton. We prefer to use the term ``coupled map lattice''
because the cell state is a real number, not an integer.
\end{thebibliography}
\end{document}